\documentstyle[sprocl]{article}

\input{epsf}

\def\NPB#1#2#3{{\em Nucl.{\,}Phys.{\,}}{\bf B{#1}}, {#2} ({#3})} 
\def\PLB#1#2#3{{\em Phys.{\,}Lett.{\,}}{\bf {#1}B}, {#2} ({#3})} 
\def\PRL#1#2#3{{\em Phys.{\,}Rev.{\,}Lett.{\,}}{\bf  {#1}}, {#2} ({#3})} 
\def\PRD#1#2#3{{\em Phys.{\,}Rev.{\,}}{\bf D{#1}}, {#2} ({#3})} 
\def\PR#1#2#3{{\em Phys.{\,}Rep.{\,}}{\bf {#1}}, {#2} ({#3})}

\def\be{\begin{equation}}
\def\ee{\end{equation}}
\def\bea{\begin{eqnarray}}
\def\eea{\end{eqnarray}}

\def\thetaeff{\theta_{eff}}
\def\scriptM{{\cal M}}

\newcommand{\bref}[1]{(\ref{#1})} 
\newcommand{\ct}[1]{\,\cite{#1}}

\begin{document}

\begin{flushright}
{CCNY-HEP-98/7 \\ }
{hep-lat/9810044 \\ }
{\hfill October 1998 \\ }
\end{flushright}
\vglue 0.8cm

\title{On the Analog Strong CP Problem in the $CP^{N-1}$ Models{\footnote{To 
appear in the proceedings of PASCOS 98.}}}

\author{Stuart Alan Samuel}

\address{Physics Department, Columbia University \\ New York, 
NY 10027, USA \\ E-mail: samuel@cuphyc.phys.columbia.edu}

\address{Department of Physics, City College 
of New York \\ 138th Street and Convent Avenue \\ New York, 
NY 10031, USA \\ E-mail: samuel@scisun.sci.ccny.cuny.edu}


\maketitle\abstracts{ Addressed is the question of 
whether a natural mechanism exists to resolve 
the strong CP problem. 
The analogous issue for the two-dimensional $CP^{N-1}$ models 
is analyzed using computer simulations.}

\section{Introduction}
Non-perturbative effects 
can introduce CP violation into the strong interactions. 
Since no such violations of this discrete symmetry 
have been observed, 
an avoidance mechanism must be operative. 
Although many proposals\ct{kim87,cheng88,mohapatra92} 
have been suggested to resolve 
this issue,
the strong CP problem remains 
one of the outstanding low-energy mysteries 
of the standard model of particle physics. 

A four-dimensional Yang-Mills theory 
has instantons. 
They represent tunnelling 
between different classical $n$-vacua, 
where $n$ is an integer characterizing 
a topological winding number. 
The true quantum vacua are believed 
to be linear combinations of the $n$-vacua 
weighted by 
$\exp (i n \theta )$.\ct{cdg76,jr76} 
Alternatively, on can add to the standard Yang-Mills action 
the term $\theta Q$, 
where 
$
 Q = g^2 / (32 \pi^2) \int d^{4} x  F_{\mu \nu}^a 
      \tilde F^{\mu \nu}_a (x)  
$
is the total instanton number. 
The resulting theory explicitly breaks parity, 
time reversal invariance and CP 
unless $\theta$ equals $0$ or $\pi$. 
When $\theta$ equals $\pi$, 
CP is believed to be spontaneously broken. 

When quarks are included in the analysis, 
the question of CP violation is more complicated. 
An overall phase in the quark mass matrix $\scriptM$ 
can be eliminated through an axial $U_A(1)$ transformation, 
but this produces an extra contribution 
to $\theta$ of $ArgÊ\ detÊ\scriptM$. 
The relevant parameter 
is thus $\thetaeff = \theta + ArgÊ\ detÊ\scriptM$. 
In QCD, strong CP violation arises 
unless $\thetaeff = 0$. {}From measurements 
on the neutron's electric dipole moment, 
a bound of $\thetaeff < 10^{-9}$ 
is obtained.\ct{baluni79,cdvw79} 
The strong CP problem can be stated as follows. 
Why is $\thetaeff < 10^{-9}$ when $\thetaeff$ 
is a expected to be a parameter of order unity? 
In other words, $\theta$ must be tuned 
to cancel the contribution to $ArgÊ\ detÊ\scriptM$ 
to one part in a billion. 
If this is adjusted by hand at the tree level, 
loop effects ruin the cancellation. 
Various proposals have been made 
to resolve this problem, 
of which the most elegant is 
the Pecci-Quinn mechanism.\ct{pq77} 
It automatically sets $\thetaeff$ to zero. 

A number of years ago, I raised the question 
of whether the strong CP problem 
in pure Yang-Mills theories could be solved 
naturally.\ct{samuel92} 
One should realize that it is strong-coupling dynamics 
that is most relevant to the issue of $\theta$ dependence. 
Why is this so? 
Long-distance physics always 
determines vacuum structure. 
For example, local fluctuations in magnetic moments 
do not determine whether an iron bar magnetizes. 
A microscopic region may look disorderd 
but the overall system can be ordered, or vice-versa. 
Since $\theta$-vacua are at the heart of the CP problem, 
the existence or non-existence 
of long-distance correlations is crucial. 
Due to our current limited understanding 
of long-distance physics and strong-coupling dynamics 
in Yang-Mills theory, 
perhaps some subtle effect 
concerning the strong CP problem has been overlooked. 

There exists a toy model 
in which the strong CP problem 
is dynamically resolved. 
It is the $(2+1)$-dimensional Georgi-Glashow model 
that was analyzed by Polyakov. 
It exhibits a natural relaxation mechanism 
by producing its own axion-like field 
to eliminate the $\theta$ dependence. 

The model has instantons: 
They are the 't Hooft-Polyakov monopoles. 
Polyakov showed that these monopoles 
lead to confinement and 
produce a mass for the unbroken vector boson. 
One can repeat Polyakov's analysis\ct{polyakov77} 
in the presence of a $\theta$ term. 
To the action, one adds $\theta Q$, 
where $Q$ is the net monopole number. 
The dilute gas approximation gives 
\be
 Z_\theta = 
  {{1} \over {\cal N}} \int {\cal D} \phi 
    \exp \left[ - {{1} \over {2}}  \int d x 
    \partial^\mu \phi \partial_\mu \phi (x) +
         2 \lambda \cos \left( \phi(x)/ \sqrt{\mu} +\theta \right)
         \right]
\quad  
\label{eq1} 
\ee
for the partition function, 
where $\lambda$ and $\mu$ are parameters. 
Here $\phi$, a field that is dynamically generated, 
reproduces the Coulombic interactions between monopoles. 
Since the shift $\phi \to \phi - \sqrt{\mu} \theta$, 
eliminates the $\theta$ dependence in $Z_\theta$, 
the $(2+1)$-dimensional analog 
of the strong CP problem is resolved. 

There is a physical reason 
why the $\theta$ dependence goes away. 
Three-dimensional Coulombic gas systems 
are overall neutral. 
If $Q = 0$, then the term $\theta \times$(monopole number) 
has no effect when added to the langragian. 

But why is the three-dimensional Coulombic gas system neutral? 
The answer is due to the attractive $1/r$ potential 
between monopoles and anti-monopoles. 
The force between the monopoles and the anti-monopole 
is sufficiently strong so as to render the gas neutral. 

Once one realizes this, 
it is easy to establish criteria 
for when the $\theta$ dependence in a theory 
naturally goes away.\ct{samuel92}  
Let $V(x)$ be the anti-instanton--instanton interaction potential 
in a $d$-dimensional system computed 
in the absence of any other instantons 
and let $R$ be the distance 
between the instanton and the anti-instanton. 
\be 
 {\rm If} \ V(R) \sim 1/R^n \ {\rm with}\ n < d 
   \ {\rm then\ there\ is\ no}\ \theta \ {\rm dependence}  
\quad . 
\label{eq2} 
\ee
For four-dimensional Yang-Mills theories, 
there is no strong CP problem 
if the instanton is attracted to the anti-instanton 
through a potential stronger that $1/R^4$ for $R$ large. 
Because of our ignorance of the long-distance behavior 
of Yang-Mills theories, it is not known 
whether the criterion in Eq.\ \bref{eq2} is satisfied. 

\section{Computer Results of Schierholz et al 
for $CP^{N-1}$ Models}

During the past few years, 
G.\ Schierholz and co-workers have performed  
some intriguing Monte Carlo studies 
for the two-dimensional  
$CP^{N-1}$ models in the presence of 
a $\theta$ term.\ct{os94,schierholz94}  
These models have several features 
in common with $d=4$ Yang-Mills theories: 
They have confinement and instantons. 

The free-energy difference per unit volume 
$f(\theta)$ is determined from  
\be 
  \exp (-Vf(\theta)) = Z(\theta )/Z(0)
\quad , 
\label{eq3} 
\ee
where $V$ is the volume of the system and $Z(\theta )$ 
is the partition function with the two-dimensional $\theta$ term  
of $\theta/(2 \pi) \int d^2 x F_{01}$. 

As a function of $\theta$, the free-energy 
obtained from Monte Carlo simulations 
exhibited a flattening 
behavior\ct{os94,schierholz94} 
that is reproduced well by 
\be
  f =
    \left\{ \begin{array}{ll} 
    a\left( \beta  \right) \theta^2 
       \quad &\theta < \theta_c \\ 
    c\left( \beta  \right) \quad \, \quad &\theta > \theta_c   
            \end{array} 
     \right. 
\quad . 
\label{eq4} 
\ee
In other words, $f$ rises quadratically 
with $\theta$, 
turns over and then displays no dependence 
on $\theta$ beyond a critical value $\theta_c$. 
For two-dimensional confining systems, 
the string-tension $\sigma$ for 
a particle of charge $e$ can be computed from 
\be 
  \sigma (e,\theta) = f(\theta+2 \pi e) - f(\theta)
\quad . 
\label{eq5} 
\ee
Thus the behavior in Eq.\ \bref{eq4} implies that 
confinement is lost for $\theta > \theta_c$ 
for sufficiently small charges $e$. 
Furthermore, the Monte Carlo simulations 
suggested that $\theta_c(\beta )$ 
goes to zero as the continuum limit is taken. 
Taken together, the above statements imply 
that a confining continuum $CP^{N-1}$ theory 
would have to have $\theta$ adjusted to zero. 
The conclusion is that the requirement of confinement 
necessitates a solution 
of the two-dimensional analog strong CP problem 
in the $CP^{N-1}$ models.\ct{os94,schierholz94}

By the way, a flattening behavior 
in $f(\theta )$ was also seen 
for the $d=4$ $SU(2)$ lattice gauge theory 
in subsequent simulations performed 
by G.\ Schierholz.\ct{schierholz95} 
In $d=4$, however, 
this does not imply the lost of confinement 
so that the intepretation of this result is not so clear. 

Inspired by Schierholz's work, 
Jan Plefka and I decided to perform our own studies. 
We were particularly intereseted 
in whether the ``deconfinement effect'' 
could be related to a natural relaxation method.

\section{General Discussion of Lattice Measurements 
of the Free-Energy for a System with a $\theta$ Term}

It is not easy to simulate a system with a $\theta$ term. 
This is because one usually uses the $\theta = 0$ vacuum 
to analyzed the $\theta \ne 0$ vacua. 
If ``barrier penetration'' effects are strong, 
poor results can ensue. 

The free-energy difference $f(\theta)$ can be computed from 
\be
  \exp (-V f(\theta)) = \sum_Q P(Q) \exp (i\theta Q)
\quad . 
\label{eq6} 
\ee 
Here, $P(Q)$ is the probability that the system 
has topological charge $Q$. 
Let $N_{MC}(Q)$ be the number of times 
that Monte Carlo configurations with topological charge $Q$ 
are generated.  
Then
\be
  P_{MC}(Q) \equiv  
  { { N_{MC}(Q) } \over { \sum_{Q^\prime} N_{MC}(Q^\prime) } }
\quad  
\label{eq7} 
\ee 
provides a Monte Carlo estimate $P_{MC}$ of $P(Q)$. 
Using Eq.\ \bref{eq6}, 
a Monte Carlo value $f_{MC}(\theta)$
of $f(\theta )$ can be obtained from 
\be
 \exp (-Vf_{MC}(\theta )) = \sum_Q P_{MC}(Q) \exp (i\theta Q)
\quad  .
\label{eq8} 
\ee
This method of computing $f(\theta)$ 
can lead to an anomalous flattening effect. 
Suppose there is a dominant statistical error 
in the $Q=0$ sector 
and that the probability of having a $Q=0$ configuration 
is over estimated: $P_{MC}(0) > P(0)$. 
Then one can show\ct{ps97b} that 
\be
  f_{MC} \approx
    \left\{ \begin{array}{ll} 
    f (\theta ) 
       \quad &\theta < \theta_b \\ 
     f ( \theta_b )  \quad \, \quad &\theta > \theta_b   
            \end{array} 
     \right. 
\quad ,  
\label{eq9} 
\ee
where $\delta P(0) = P_{MC}(0) - P(0)$ and 
\be
   f ( \theta_b ) \approx { {1} \over {V} } | \log | \delta P(0) | | 
\quad .    
\label{eq10} 
\ee 

Here are the key points 
(see reference [14]):

(1) When the volume is sufficiently big, 
a limiting $\theta_b$ arises.  For $\theta > \theta_b$, 
$f(\theta)$ cannot be reliably measured.

(2) If a flattening behavior in $f(\theta)$ is seen, 
one should be suspicious of the result. 
One needs to check whether $|\delta P(0)|$ 
is bigger than it should be.

(3) If a large-$V$ simulation shows a flattening effect 
for $f(\theta )$, 
but a smaller-$V$ simulation does not, 
then one should probably trust the smaller-$V$ result.

In refs.[11,12],
no flattening effect was seen in simulations 
with small volumes. 
Point (3) says that one should trust 
the smaller volume studies. 
If true, the $CP^{N-1}$ models 
do not undergo a deconfining phase transition.

\section{Monte Carlo Results 
for the Exactly Solvable $d=2$ $U(1)$ Gauge Theory}

One way to decide the question 
of anomalous flattening behavior 
is to study an exactly solvable system. 
The $d=2$ $U(1)$ gauge theory is such a case. 
The action is given by 
\be
    S^{U(1)} = \beta \sum_p 
             \left( { U_p + U_p^{*}  } \right) 
\quad , 
\label{eq11}
\ee 
where $U_p$ is the product of the $U(1)$ link phases 
around a plaquette $p$ 
and $\beta$ is the inverse coupling. 

The lattice analog of the continuum $\theta$-term action 
$i { {\theta} / {2 \pi} } \int d^2 x F_{01}$ is 
\be 
  S_{\theta \ {\rm term}} = 
  {{ \theta } \over {2\pi} } \sum_p 
   \log \left( { U_p } \right) 
\quad . 
\label{eq12} 
\ee 
The exact analytic result for the partition function 
governed by the sum of the action 
in Eqs.\ \bref{eq11} and \bref{eq12} 
has been computed 
in ref.[15]. 

When Jan Plefka and I performed 
Monte Carlo simulations of this system, 
we sometimes found a flattening behavior 
in the free energy that disagreed 
with exact analytic results.\ct{ps97b} 
Figure 1 illustrates this.  

\begin{figure}[t]
\epsfxsize=5in
\epsfbox{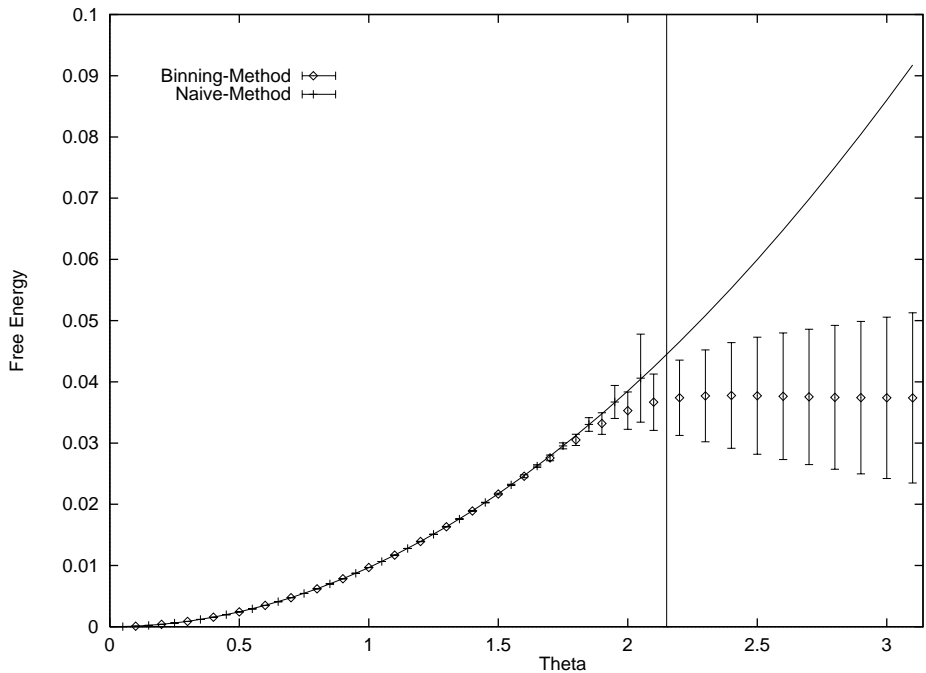}
\caption{The Free Energy of the U(1) Gauge Model 
Versus $\theta$. The solid line is 
the exact result.  \label{figure1}}
\end{figure}

Noteworthy features are: 

(1) A barrier $\theta_b$ arises. 

(2)  A short calculation verifies that $\theta_b$ 
is the value expected from Eq.\ \bref{eq10}. 

(3) The flattening effect is due to $P_{MC}(0) > P(0)$.

It turns out that
$\theta_b$ moves to smaller values as $V$ increases.
For a $7 \times 7$ lattice, 
the free energy agreed with exact analytic results 
for all $f(\theta ) $.

\section{Monte Carlo Results for a  $CP^{N-1}$ Model}

We also decided to simulate the $CP^{3}$ model 
for a lattice action involving an auxiliary $U(1)$ gauge field. 
The advantage of using this latticization is 
that analytic strong coupling series are 
available.\ct{seiberg84,ps97a} 
The action is given by 
\be 
  S = \beta N \sum_{x,\Delta} 
 \left( { 
    z_{x}^\ast \cdot z_{x+\Delta} 
    U \left({  x, x+\Delta } \right) + 
    z_{x} \cdot z^\ast_{x+\Delta} 
    U^\ast \left({  x, x+\Delta } \right)
        } \right) 
\quad , 
\label{eq13} 
\ee 
where the complex scalar fields $z_{x}^i$ 
satisfy $\sum_{i=1}^N z_{xi}^\ast z_{x}^i =1 $ 
and $N=4$ for the $CP^3$ model. 
Here,   
the sum is over 
unit shifts $\Delta$
in the $d$ positive directions, 
where $d=2$. 
The field 
$U \left({  x, x + \Delta } \right)$ 
is a phase associated with the link between 
$x$ and $x + \Delta$:  
it is the $U(1)$ auxiliary gauge field.  
Eq.\ \bref{eq13} 
and $S_{\theta}$ 
of Eq.\ \bref{eq12} are added 
to obtain the full action.  
This lattice action differs from the one used 
in refs.[11,12] 
but is equally good.

When we performed Monte Carlo simulations, 
again, spurious flattening behavior 
in the free energy sometimes occurred. 
Figure 2 is an example. 

\begin{figure}[t]
\epsfxsize=5in
\epsfbox{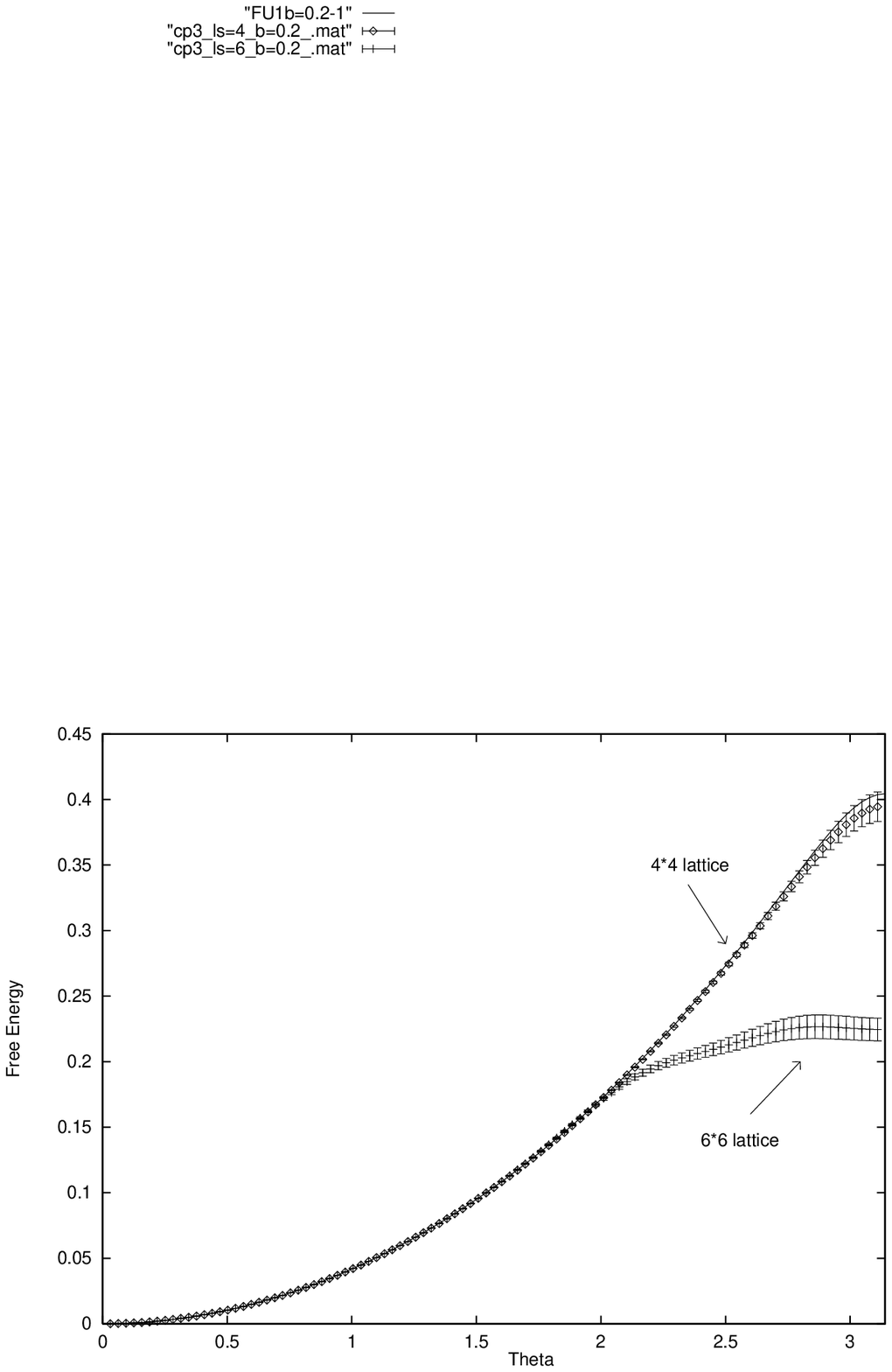}
\caption{The $CP^3$ Free Energy Versus $\theta$ 
at $\beta=0.2$. The solid line is the strong coupling 
series.  \label{figure2}}
\end{figure}

\section{Conclusion}

Here are the main points: 

(1) Subtleties arise when simulating systems 
with a $\theta$ term.  
The basic difficulty is that one is trying 
to simulate one vacuum (a $\theta \ne 0$ vacuum) 
using another vacuum ($\theta =0$ vacuum).  
For good Monte Carlo results, barrier penetration 
needs to be suppressed.  
When the volume of the system is large, 
the barrier penetration becomes 
sufficiently severe that reliable results cannot be obtained, 
particularly for $\theta$ near $\pi$. 

(2) Because current simulation methods tend 
to have the biggest error in the measurement 
for the $Q=0$ probability, 
Monte Carlo results can generate 
an anomalous flattening behavior 
in the free-energy for $\theta$ large. 

(3) Reliable results for the free-energy appear 
to be obtainable throughout the entire $\theta$ region 
as long as the volume is keep sufficiently small.

(4) The ideas and results of this presentation 
suggest that 
the $CP^{N-1}$ model with a $\theta$ term 
does not undergo a deconfining phase transition.

(5) In short, it is unlikely that 
the solution of the strong CP problem 
as suggested by previous numerical studies 
in the $CP^3$ model actually works.

\section*{Acknowledgments}
I thank Jan Plefka 
for his contributions to this work. 
He played an essential role 
in the developing strong coupling methods 
for the $CP^{N-1}$ models 
as well as in the computer studies of $CP^3$. 
I thank the Max-Planck Institute 
in Munich, Germany -- 
much of the computing 
was carried out on MPI work stations.   
This research is supported in part 
by the PSC Board of Higher Education at CUNY and   
by the National Science Foundation under the grant  
(PHY-9605198).


\section*{References}
\begin{thebibliography}{99}

\bibitem{kim87}
J.\,E.\,Kim, 
\PR{150}{1}{1987}.

\bibitem{cheng88} 
H.\,Y.\,Cheng, 
\PR{158}{1}{1988}.

\bibitem{mohapatra92}
See Chapter 4 of R.\,N.\,Mohapatra, 
{\it Unification and Supersymmetry}, 
(Springer-Verlag, New York, 1992) and references therein.

\bibitem{cdg76} 
C.\,Callan, R.\,Dashen and D.\,Gross, 
\PLB{63}{334}{1976}. 

\bibitem{jr76} 
R.\,Jackiw and C.\,Rebbi, 
\PRL{37}{172}{1976}. 

\bibitem{baluni79} 
V.\,Baluni, 
\PRD{19}{2227}{1979}. 

\bibitem{cdvw79} 
R.\,Crewther, P.\,Di Vecchia, 
G.\,Veneziano and E.\,Witten, \\ 
\PLB{88}{123}{1979}. 

\bibitem{pq77} 
R.\,Peccei and H.\,Quinn, 
\PRL{38}{1440}{1977}. 

\bibitem{samuel92} 
S.\,Samuel, 
Mod.{\,}Phys.{\,}Lett.{\,} {\bf A7} ({1992}) {2007}.  

\bibitem{polyakov77}
A.\,Polyakov, 
\NPB{120}{429}{1977}. 

\bibitem{os94}  
S.\,Olejnik and G.\,Schierholz, 
{\em Nucl.\,Phys.\,B (Proc.\,Suppl.)}{ \bf 34} (1994) 709. 

\bibitem{schierholz94}  
G.\,Schierholz,
{\em Nucl.\,Phys.\,B (Proc.\,Suppl.)}{ \bf 37A} (1994) 203.  

\bibitem{schierholz95} 
G.\,Schierholz,
{\em Nucl.\,Phys.\,B (Proc.\,Suppl.)}{ \bf 42} (1995) 270. 

\bibitem{ps97b} 
J.\,Plefka and S.\,Samuel, 
{\it Monte Carlo Studies of Two-Dimensional Systems 
with a $\theta$ Term}, 
\PRD{56}{44}{1997}, hep-lat 9704016.

\bibitem{wiese89} 
U.-J.\,Wiese, 
\NPB{318}{153}{1989}. 

\bibitem{seiberg84} 
N.\,Seiberg, 
\PRL{53}{637}{1984}.  

\bibitem{ps97a} 
J.\,Plefka and S.\,Samuel, 
\PRD{55}{3966}{1997}, hep-lat 9612004.


\end{thebibliography}
\end{document}